\documentstyle[aps,epsf,twocolumn]{revtex}
\begin{document}
\title{Maxwell Equations and Irreversibility}\author{Mario Liu\\ Institut 
f\"ur Theoretische Physik, Universit\"at Hannover,\\30167 Hannover, Germany, 
EC}\date{\today}\maketitle

\begin{abstract}Two questions connected to the macroscopic Maxwell 
equations are addressed: First, which form do they assume in the 
hydrodynamic regime, for low frequencies, strong dissipation and arbitrary 
field strengths. Second, what  does this tell us about irreversibility and 
coarse-grained description. \end{abstract}

\tableofcontents
\section{Introduction}
You are probably ambivalent toward the article you have just started to 
read, or even harbor dark suspicions about it. These feelings are admittedly 
hard to avoid when encountering someone purporting to reveal news about the 
centenarian Maxwell equations. Afterall, we know all there is to know about 
the Maxwell equations, do we not? Well, no, we do not. Despite the 
impression we came away with from various courses on electrodynamics, there 
are large gaps in our understanding on this subject, and the most glaring 
ones deal with the connection between electrodynamics and thermodynamics. 
These two independently developed classical areas of physics apparently do 
not mix well: Jackson (1975) does not mention entropy in his classic book at 
all, while Callen (1985) has eliminated the (partly erroneous) chapter on 
magnetic and electric system in the second edition of his definitive work. 

Nevertheless, it must be sensible to ask questions such as when an 
electromagnetic field configuration is in equilibrium, and how this 
equilibrium state is arrived at. Equilibrium fields are the ones that maximize the 
entropy, is there a simple equation expressing this property? Equilibrium fields 
are always static, but is it possible, in dielectrics,  for a static field to be 
off-equilibrium? These are some of the questions answered below, and we go 
on from these answers to understand how field dissipation can be accounted 
for in dynamic  situations, for a general system of nonlinear constitutive 
relations -- where postulating imaginary parts for the permeabilities 
$\varepsilon$ and $\mu$ fails to work. We also go on to understand what the 
electromagnetic force is that a polarizable and magnetizable body feels -- 
both in and off equilibrium. Fortunately, all these results are fairly 
simple and universal,  and in fact quite suitable for being introduced into 
a university course on ``electrodynamics in continua''. 

In section \ref{reso}, I shall summarize the state of the art of the 
Maxwell equations as it is conventionally treated, and point out the extent 
of securely understood physics. Starting with section \ref{thermo}, a 
different approach, that of the thermo- and hydrodynamic theory, is 
introduced. The hydrodynamic theory, the response of dense and 
dissipative systems exposed to slowly varying external fields of nonlinear 
strength, is discussed in section \ref{hyd}. Interspersed between these 
considerations, I shall often stop to deliberate over two fundamental 
aspects of macroscopic theories ---  coarse-grained description and 
irreversibility --- starting right at the beginning of the next section, but 
mainly in section \ref{cf}. The difference between the hydrodynamic and the 
linear response theory is dwelt on in section \ref{linres}, with some 
surprising and instructive results. Finally, in section \ref{force}, the old 
but confusing subject about the electromagnetic force in a coarse-grained 
description is considered and clarified.  

{\large\bf[}Arguments that are not usually reproduced in a seminar, only 
invoked when the appropriate question is raised, are given -- as here -- in 
large square brackets. They may be skipped on first reading.{\large\bf]}

\section{Less Accurate Is More Difficult?}\label{reso}
The actual difference between the electrodynamics in vacuum and continua is 
one of accuracy, better: one of resolution. Denoting the grain size of 
descriptive grid as $\xi$, and the average distance between the charge 
carriers as $d$, the vacuum Maxwell equations as a high resolution theory 
are valid for $d\gg\xi$. Since the input, the density of charge and current, 
is in principle arbitrarily accurate, we are able to calculate the fields 
${\bf e}$ and ${\bf b}$ to any desired resolution. 

Their contribution to the energy, and the Lorentz force are, respectively  

\begin{equation}\label{1}
u^{\rm em}=\textstyle\frac{1}{2}(\varepsilon_0e^2+b^2/\mu_0),\quad
{\bf f}=\varrho_e({\bf e}+\mbox{\boldmath$v$}\times{\bf b}).  
\end{equation}
In conjunction with the Newtonian equation of motion, these two expressions 
account for the feedback, for how the field affects the motion of the 
material. So we have at our disposal a closed theory --- a classical one 
with a known quantum mechanical generalization  --- that is conceptually 
simple yet technically intractable for dense systems.

The circumstances are reversed if we check our ambition and seek 
enlightment from a low-resolution theory, $d\ll\xi$: The technical  
difficulties are greatly reduced, but conceptually we enter murky waters. 

Because the  density of charge $\langle\varrho_e\rangle$ and current  
$\langle j_e\rangle=\langle\varrho _ev\rangle$, are now spatially averaged 
quantities, we must deal with hidden charges and currents, making it 
necessary to consider four instead of two fields: ${\bf D}$, ${\bf B}$, 
${\bf E}^M$, and ${\bf H}^M$ --- all coarse-grained, hence in capital 
letters.  (As will become clear soon, we need to distinguish ${\bf E}^M$, 
${\bf H}^M$ from ${\bf E}$, ${\bf H}$. The former two are the usual fields 
as defined by the Maxwell equations; the latter two will be introduced later.) 
Given constitutive relations, linear if the field is sufficiently weak, 

\begin{equation}\label{3}
{\bf D}=(\varepsilon'+i\varepsilon''){\bf E}^M,\quad{\bf B}= 
(\mu'+i\mu''){\bf H}^M, \end{equation} we are again able to calculate the 
field from the source. The real parts of the permeabilities, $\varepsilon'$ 
and $\mu'$, account for the reactive responses such as the oscillatory 
motion of hidden charges; the imaginary parts, $\varepsilon''$ and $\mu''$, 
parametrize dissipation and absorption. If a field is  stronger, it is 
customary to take the corresponding permeability again as a function of the 
field. But one goes beyond  linear response only at the price of loosing all 
the simple relations, especially the identification of the imaginary part 
with dissipation. 

Deplorable as this is, the problems are worse for the feedback, the effect 
of the field on the motion of the material.  The reason is that both the 
electromagnetic energy and the Lorentz force, Eqs(\ref{1}), are nonlinear. 
And the knowledge of the coarse-grained quantities $\langle\varrho_e\rangle$, 
$\langle{\bf j}_e\rangle$, ${\bf E}^M=\langle{\bf e}\rangle$, ${\bf 
B}=\langle {\bf b}\rangle\dots$  is quite useless if we need to know the 
values of $\langle\varrho_e{\bf e}\rangle$, or 
$\langle\varrho_e\mbox{\boldmath$v$}\times{\bf b}\rangle$.  Nevertheless, 
bold souls, without much ado, simply write

\begin{equation}\label{5}
 \rho{\textstyle\frac{\rm d}{{\rm d}t}} v_i +\nabla_i P+ \nabla_j\pi^D_{ij} 
= \left[\langle\varrho_e\rangle{\bf E}^M+\langle{\bf j_e}\rangle\times{\bf 
B}\right]_i.\end{equation}
In the absence of fields, the right side of Eq(\ref{5}) is 
zero, and what remains is the Navier-Stokes equation, an expression of 
momentum conservation in the low-resolution, hydrodynamic 
physics: Both the pressure $P$, a quintessentially thermodynamic quantity, 
and the viscous stress tensor $\pi^D_{ij}$, a dissipative term, presume an 
infinitesimal volume element that contain enough particles to form a system 
in local equilibrium. This volume element is nothing but the descriptive 
grain, of the size $\xi^3$, we therefore have $d\ll\xi$.  Now, any theory, 
and each equation, must consistently have a unique resolution. So the 
sources and fields on the right hand side must not be of high resolution, 
but it is not at all clear that the expression as written, in low-resolution 
quantities, is correct. Needless to say, anyone employing this fairly 
popular equation, or a variant of it, bears the burden of proof for its 
validity. {\large\bf[}An example of a system in which Eq(\ref{5}) does hold 
is a weakly dissociated gas, where the density of charge carriers  is much 
lower than the density of neutral particles. Assuming negligible dipole 
moments for the latter, we have two interparticle distances, $d_n$ for 
neutral particles,  and $d_e$ for the charge carriers. Then the resolution 
of this equation may be chosen as $d_n\ll\xi\ll d_e$, such that it is of 
low-resolution for the neutral particles, yet of high resolution for the 
charge carriers.{\large\bf]}

Actually, things are not as bleak as they seem right now, and a few lesser 
known results do considerably brighten up the prospect of formulating the 
feedback, and closing the low-resolution Maxwell theory. These results are 
(i) the thermodynamics of the electromagnetic field and (ii) the expressions 
for the energy and stress tensor at finite frequencies, albeit without 
dissipation. They will be briefly outlined in the next section and are taken 
from the one good book on this subject, Volume VIII of Landau and Lifshitz 
(1984), referred to below as LL8; see also Kentwell and Jones (1987) for a 
nice review of the state of the art. The thermodynamic 
considerations as presented, however, contain enough generalization and shifts 
in interpretation, from the unfortunately brief treatment in \S18 of LL8, that 
I most probably have to bear the blame if you should find faults 
in any statements here. 

\section{The Thermodynamics of Fields}\label{thermo}
We start from the energy density $u$ in the system's rest frame, as a 
function of the entropy density $s$, mass density $\rho$, and the two fields 
${\bf D}$ and ${\bf B}$, 

\begin{equation}\label{7}
{\rm d}u=T{\rm d}s+\mu{\rm d}\rho+{\bf E}\cdot {\rm d}{\bf 
D}+{\bf H}\cdot {\rm d}{\bf B}.
\end{equation}
Being defined as $\partial u/\partial{\bf D}$ and $\partial u/\partial 
{\bf B}$, respectively, ${\bf E}$ and ${\bf H}$ are (like $T\equiv \partial 
u/\partial s$) functions of all the thermodynamic variables. For weak 
fields (and barring ferroelectricity or ferromagnetism), an expansion yields 

\begin{equation}\label{7a}
{\bf E}={\bf D}/\bar\varepsilon,\quad {\bf H}={\bf B}/\bar\mu, 
\end{equation}
 where $\bar\varepsilon$ and $\bar\mu$ may be functions of temperature and 
density, but not the frequency. 

Maximizing the entropy $S=\int\!dV s$  for a stationary dielectric medium, 
with the constraints of  constant energy, constant mass, and the validity of 
the two non-temporal Maxwell equations, $\nabla\cdot{\bf D}= 
\langle\varrho_e\rangle$ and $\nabla\cdot {\bf B}=0$, the resulting Euler 
equations are 

\begin{eqnarray}\label{8}
 \nabla T=0,\quad \nabla\mu=0, \\\label{9}\nabla\times{\bf 
E}=0,\quad\nabla\times{\bf H}=0.\end{eqnarray}
The entropy is maximal, and the system in equilibrium, only if these 
equations are satisfied. {\large\bf[}As an illustration, keep only $\bf D$ 
as a variable and minimize $U=\int u$. This leads to $\int{\bf 
E}\!\cdot\!\delta{\bf D}- \Phi({\bf r})  \delta(\nabla\!\cdot\!{\bf D} - 
\langle\varrho_e\rangle) =0$, with $\Phi({\bf r})$ a Lagrange parameter. A 
partial integration and the fact that $\delta{\bf D}$ is arbitrary yield 
${\bf E}=-\nabla\Phi $, or the first of Eqs(\ref{9}). Circumstances are 
modified if a small but finite conductivity eventually allows the charge to 
move, rendering $\langle\varrho_e\rangle$ no longer constant, only 
$\int\langle\varrho_e\rangle$. Then the entropy can be further increased, 
and becomes maximal for ${\bf E}=0$.{\large\bf]}

The two Eqs(\ref{9}) are a surprise, as they amount to a thermodynamic 
derivation of the static Maxwell equations. Irrespective of what specific 
form $u$ --- and hence ${\bf E}$ and ${\bf H}$ --- assume, Eqs(\ref{9}) hold. 
{\large\bf[}To see that ${\bf E}$ and ${\bf H}$ are indeed the measured 
fields, we remind ourselves that the two variables ${\bf D}$ and ${\bf B}$ must 
reduce to ${\bf e}$ and ${\bf b}$ in vacuum, or a rarefied gas. A comparison 
of the  two energy expressions, Eq(\ref{7}) and (\ref{1}) then compels ${\bf 
E}$ and ${\bf H}$  to do the same, reduce to ${\bf e}$ and ${\bf b}$. The 
usual boundary conditions then imply that these four fields of a dense 
system are the ones that will be appropriately continued into an adjacent 
vacuum, where field measurements are easily carried out.{\large\bf]}

This thermodynamic consideration -- in conjunction with its hydrodynamic 
generalization in section \ref{hyd} -- pries open a new door to understand the 
macroscopic Maxwell equations. It possesses all the advantages of a thermo- 
and hydrodynamic theory, being general, independent of microscopic 
interactions, and valid for arbitrary field strength. At the same time, it 
is irreversible,  and intrinsically of low resolution. 

In the presence of fields, particles can be moved and accelerated, and the 
momentum density $\rho\mbox{\boldmath$v$}$ of the material is not a 
conserved quantity. However, because the Hamiltonian remains invariant under 
translations that include both the material and the charges producing the 
field, the total momentum of material and field is still conserved, 

\begin{equation}\label{11}
\mbox{\boldmath$\dot g$}^{\rm tot}+\nabla_j\Pi_{ij}=0.
\end{equation}
This total momentum density is 
\begin{equation}\label{12}
\mbox{\boldmath$g$}^{\rm tot}=\rho\mbox{\boldmath$v$}+{\bf E}\times{\bf 
H}/c^2, \end{equation}
because (i) the main contribution in the relativistic, total energy current 
is the sum of the rest mass motion and the 
Poynting vector, ${\bf Q}^{\rm rel}=[\rho c^2+{\cal O}(u)] 
\mbox{\boldmath$v$}+{\bf E}\times{\bf H}$; and (ii) the 4-energy-momentum 
tensor is symmetric, ${\bf Q}^{\rm rel}/c^2= \mbox{\boldmath $g$}^{\rm tot} 
$. Despite controversies about the form of $\mbox{\boldmath$g$}^{\rm tot}$ 
that refuses to die down, it appears difficult to circumvent this simple and 
fundamental argument.  

On the other hand, for the gist of this paper, it is not important which 
form $\mbox{ \boldmath $g$}^{\rm tot}$ assumes, as long as it is a definite 
one. The point is, given the expressions for $g_i^{\rm tot}$ and $\Pi_{ij}$, 
the acceleration $\frac{\partial} {\partial t} (\rho v_i)$ may be 
calculated. Therefore, the knowledge of these two expressions contains that 
about  the coarse-grained Lorentz force.

The flux $\Pi_{ij}$ is, in equilibrium and at vanishing velocity, the 
Maxwell tensor. It may be derived by considering the change in the total 
energy, of a system containing charges, when a certain portion of its 
surface is moved, see \S15 of LL8. The result is a longish expression 
containing only thermodynamic and conjugate variables, 

\begin{eqnarray}\label{MS}
\Pi_{ij}&=&(Ts+\mu\rho+E_iD_i+H_iB_i -u)\delta_{ij}\\
&-& [E_iD_j+H_iB_j+(i\leftrightarrow j)]/2.\nonumber
\end{eqnarray}
 The expression for the flux at finite frequencies is a less trivial matter 
and presupposes the corresponding expression for the energy. Assuming linear 
constitutive relations, lack of dissipation ($\varepsilon'',\mu''=0$), 
quasi-monochromacy and stationarity ($v\equiv0$), Brillouin showed in 1921 
that the additional energy due to the presence of fields is 

\begin{equation}\label{13}
 {\textstyle\frac{1}{2}} \langle E^2\rangle {\rm d} (\omega\varepsilon')/ 
{\rm d}\omega+ {\textstyle\frac{1}{2}} \langle H^2\rangle {\rm d} 
(\omega\mu')/ {\rm d}\omega,\end{equation}where the average is temporal, 
over a period of oscillation. Compared to the corresponding thermodynamic 
expressions, Eqs(\ref{7}, \ref{7a}), there are two new terms 
$\sim\omega {\rm d}\varepsilon'/{\rm d} \omega$, 
$\omega {\rm d}\mu'/{\rm d} \omega$. Forty years later, Pitaevskii 
showed that under essentially the same assumptions, the stress 
tensor retains its form from equilibrium, Eq(\ref{MS}), and remarkably, 
does not contain any frequency derivatives, cf \S80, 81 of LL8. 

\section{Some New Results}\label{some}
If we draw a diagram of field strength versus frequency, with the field 
strength pointing to the right, and $\omega$ pointing upward, see Fig~1,

\vspace{1cm}
\begin{figure}[ht]
\begin{center}\leavevmode
\epsfxsize 0.4\textwidth\epsfbox{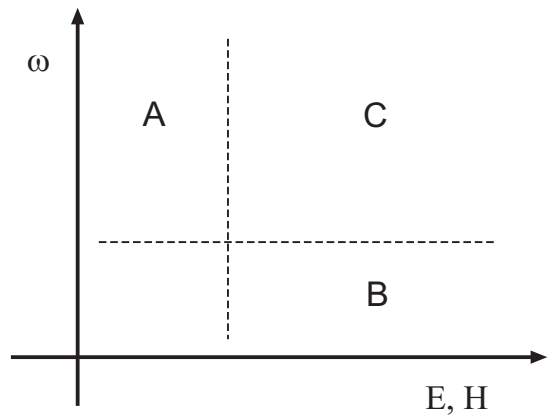}\end{center}
%\vspace{-3cm}
\caption{Range of validity}\label{fig 1}
\end{figure}\noindent

we have a vertical stripe A along the $\omega$-axis --- arbitrary frequency 
but small field strength --- that is the range of validity of the linear 
response theory, Eq(\ref{3}), while the field-axis itself depicts the space 
in which thermodynamics holds; the expressions of Brillouin and Pitaevskii 
are valid within the linear response stripe A, in isolated patches disjunct 
from the field-axis, wherever field dissipation is negligible. 
{\large\bf[}Because $\varepsilon'', \mu''$ are odd functions of $\omega$, 
and $\varepsilon', \mu'$ even ones, and because the first frequency 
dependent effects when leaving the equilibrium are linear in $\omega$, they 
belong to $\varepsilon'', \mu''$ and are dissipative. Lack of dissipation 
therefore characterizes a frequency region disjunct from 
$\omega=0$.{\large\bf]}

Having understood the thermodynamic behavior of a system, it is fairly easy 
to derive the corresponding hydrodynamic theory. It 
accounts for the same system --- condensed matter, charged or exposed to an 
external field --- that is now slightly out of equilibrium, to linear order 
in the frequency. Its range of validity is the horizontal stripe B along the 
field axis, for arbitrary field strength and small frequencies. The theory 
as derived (Liu 1993, 1994, 1995) is closed; it includes both the 
macroscopic Maxwell equations and the expression for the total momentum flux 
$\Pi_{ij}$. As terms of second order in the frequency are neglected, the 
hydrodynamic theory does not account for dispersion. 

The parameter space C beyond the two perpendicular stripes needs a theory 
that can simultaneously account for dissipation, dispersion, nonlinear 
constitutive relations and finite velocities. Although one might expect 
principal difficulties in setting up such a theory --- due to the apparent 
lack of a small parameter --- the system is in fact, up to the optical 
frequency $\sim10^{15}$Hz,  still in the realm of macroscopic physics, as 
the associated wavelength remains large compared to the atomic graininess. 
And when asking questions such as what is the force on a volume element 
exerted by a strong laser beam, if we confine our curiosity to the averaged 
force --- with a temporal resolution larger than the time needed to 
establish local equilibrium (which itself is much larger than the light's 
oscillatory period for the frequency range under consideration) --- a 
simple, universal and hydrodynamic-type theory is still possible. Better: it 
may be cogently derived, as the respective limits of small field and low 
frequency are firmly anchored. A first step toward such a theory has been successful. 
It includes the dynamics of polarization, but neglects magnetization 
(Jiang and Liu 1996). 

\section{The Hydrodynamics of Fields}\label{hyd}
Although the hydrodynamic theory of electromagnetism may appear unorthodox 
at times, it is rather elementary in its essence, and especially easy to 
comprehend by analogy. Consider a typical hydrodynamic equation, that of 
Navier-Stokes in the absence of fields, $\dot g_i+\nabla_j\Pi_{ij}=0$. The 
momentum density $g_i=\rho v_i$ is a thermodynamic variable, odd under time 
inversion. The stress tensor $\Pi_{ij}=\pi_{ij}+\pi^D_{ij}$ is the 
corresponding flux, with two parts: The reactive one is (if linearized) 
given by the pressure, $\pi_{ij}=P\delta_{ij}$, a thermodynamic derivative. 
It is even under time reversal, same as $\dot g_i$. The dissipative part, 

\[\pi^D_{ij}\sim 
v_{ij}\equiv{\textstyle\frac{1}{2}}(\nabla_iv_j+\nabla_jv_i),\] is odd and 
breaks the time inversion symmetry of the equation. If we had included the 
momentum density $g_i$ as an additional variable in the thermodynamic 
consideration above, $v_{ij}=0$ would have been added to 
Eqs(\ref{8},\ref{9}) as the respective Euler equation. If it is satisfied, 
the entropy is maximal with respect to the distribution of $g_i$. So 
understandably, if $v_{ij}\not=0$, there is a current $\sim v_{ij}$ to 
redistribute $g_i$, such that the system is pushed toward maximal entropy 
and equilibrium. This is how dissipation and irreversibility are generally 
accounted for in hydrodynamic theories. Every statement in this paragraph 
has its counterpart in the next. 

The Maxwell equations for neutral, dielectric media, ${\bf\nabla \cdot 
D}=0$, ${\bf\nabla \cdot B}= 0$, 

\begin{equation} \label{14}
{\bf\dot D}=\nabla\times{\bf H}^M,\quad{\bf \dot B}=-\nabla\times{\bf 
E}^M,   \end{equation}
impose the analogy $\mbox{\boldmath$g$}\to{\bf D,B}$ and $\Pi_{ij}\to{\bf 
H}^M$, ${\bf E}^M$:  The thermodynamic variables are ${\bf D}$ and ${\bf 
B}$, being even and odd, respectively. Eqs(\ref{14}) are their equations of 
motion, while the two non-temporal Maxwell equations are constraints. 
{\large\bf[}Eqs(\ref{14}) must have this form to ensure that the two 
constraints are satisfied at all time. And we already know ${\bf E}={\bf 
E}^M$ and ${\bf H}={\bf H}^M$ in equilibrium.{\large\bf]} The fields ${\bf 
H}^M$ and ${\bf E}^M$ appear only where fluxes do, they therefore split into 
reactive and dissipative parts, 

\begin{equation}\label{15}   
{\bf H}^M={\bf H+H}^D,\qquad {\bf E}^M={\bf E+E}^D.
\end{equation}
The reactive ones are again thermodynamic derivatives, while the 
dissipative fields ${\bf H}^D$ and ${\bf E}^D$ are proportional to 
$\nabla\times{\bf E}$ and $\nabla\times{\bf H}$, respectively, for the 
same reason as above: If these quantities are nonvanishing, the entropy is 
not maximal with respect to ${\bf D}$ and ${\bf B}$, so dissipative fields 
are generated to push ${\bf  D}$ and ${\bf B}$ toward equilibrium. For an 
isotropic system, we have

\begin{eqnarray}\label{16a}
{\bf H}^D&=&-(\alpha/\mu_0) \nabla \times{\bf E},\\ 
{\bf E}^D&=&(\beta/\varepsilon_0)\nabla\times{\bf H}+\gamma\nabla 
T,\label{16}
\end{eqnarray}
where $\gamma\nabla T$ is a cross term, similar to that producing the 
Peltier effect. Again, since $(\alpha/\mu_0) \nabla \times{\bf E}$ and 
$(\beta/\varepsilon_0) \nabla\times{\bf H}$ are of opposite parity under 
time reversal as  ${\bf\dot D}$ and ${\bf\dot B}$, respectively, they 
account for the irreversibility of the macroscopic Maxwell equations. The 
transport coefficients are essentially the relaxation time of magnetization 
and polarization, respectively,  

\begin{equation}\label{17}\alpha=\tau_M(1-\mu_0/\bar\mu), \quad 
\beta=\tau_P(1-\varepsilon_0/\bar\varepsilon). \end{equation}
{\large\bf[}This can be shown with a simple relaxation Ansatz for the 
magnetization and polarization, which to linear order in $\omega\tau_M$ and  
$\omega\tau_P$ yields the two dissipative terms of 
Eqs(\ref{16a},\ref{16}).{\large\bf]}

This concludes the brief presentation of the hydrodynamic Maxwell 
equations, the first side of the complete theory. Before we consider their 
ramifications, in section \ref{linres}, and discuss the flip side, the force 
on volume elements exerted by these fields, in section \ref{force}, there is 
a prevalent misunderstanding  concerning the coarse-graining procedure that 
we need to address first. 

\section{Coarse-Graining and Irreversibility}\label{cf}
A view point one frequently encounters in textbooks takes the 
macroscopic Maxwell equations as the spatially averaged version of the 
microscopic ones. If true, it contradicts the hydrodynamic Maxwell equations: 
Comparing ${\bf\dot b}=-\nabla\times{\bf e}\ $ to 
$\ {\bf \dot B}= -\nabla\times{\bf E}^M$, it 
seems compelling that (i) ${\bf E}^M=\langle{\bf e}\rangle$, 
and (ii) ${\bf E}^M$ must remain even under time reversal, as this property 
cannot be altered by a spatial integration. Therefore, ${\bf E}^M$ must not 
contain any odd, dissipative terms such as 
$(\beta/\varepsilon_0)\nabla\times{\bf H}$.  

Spatially averaging any microscopic equation of motion does not usually 
lead to a coarse-grained, macroscopic dynamics: An initial macrostate 
contains a large number of microstates which generically evolve into final 
microstates that belong to very different macrostates. This lack of 
uniqueness renders a macroscopic dynamics impossible to formulate, since for 
the prediction of the final macrostate one needs the knowledge of the actual 
microstate. {\large\bf[}The averaged dynamics is of course unique if the 
microscopic dynamics is strictly linear -- but any nonlinear term changes 
this qualitatively. And as emphasized above, the complete microscopic 
electrodynamics is nonlinear.{\large\bf]}

Fortunately, we are only interested in the time evolution of the field 
actually measured, not every single microscopic one, even if spatially 
averaged. So we can restore uniqueness by taking an ensemble average, the 
average of all microstates contained in a given macrostate. Quite 
frequently, especially if local equilibrium holds, this operation alters the 
time inversion property of the relevant coarse-grained field, as in the 
following example due to Onsager.

Consider a macroscopic variable $x$ that vanishes in equilibrium. Expand 
the entropy, $S=S_0-\gamma x^2$, to quantify the exponentially diminishing 
probability of higher values of $x$, and the strongly reduced number of 
microstates compatible with them. Assume that $x$ relaxes 
quasi-stationarily, ie (partial) equilibrium for a given value of $x$ 
is (compared to its slow relaxation time) established instantaneously. Then, 
given the initial value $x_0(t_0)$, the vast majority of 
microstates in this ensemble will at $t_0$ arrive in $x_0$ as their 
excursion peak while undergoing fluctuations from equilibrium, $x=0$, see Fig~2. 

\vspace{1cm}
\begin{figure}[ht]
\begin{center}\leavevmode
\epsfxsize 0.4\textwidth\epsfbox{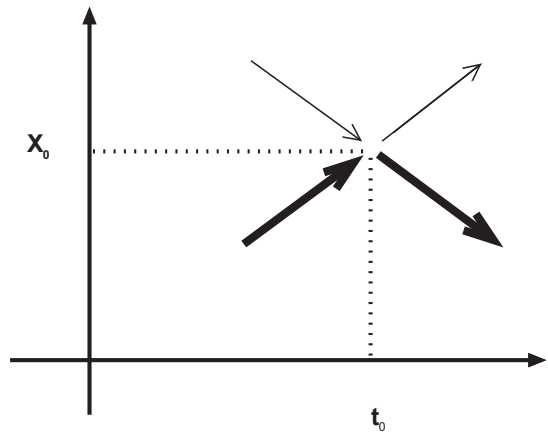}\end{center}
\caption{Furthest point of excursion}\label{fig2}
\end{figure}\noindent
A few will overshoot a moment later to attain higher values $x>x_0$ that are far 
less probable, still others will have come from higher values. But the 
majority will turn around at  $x_0$ to move back towards $x=0$. 

The ensemble average will follow suit, to arrive at $x_1$ just slightly 
smaller than $x_0$. Here, the ensemble is considerably modified by 
re-establishing equilibrium, as further microstates, with $x_1$ as 
their excursion peak, join in. As the new members greatly outnumber the old 
ones, the ensemble average now follows these to move still closer to 
equilibrium. It is this ensemble-averaging at every step, over an ever 
increasing group of microstates representing ever more probable macrostates, 
that increases the entropy, forces the macrostate towards equilibrium and 
renders the macroscopic dynamics irreversible.

In the macroscopic formulation, $\dot x=-x/\tau$, this physics is 
accounted for by the dissipative term on the right side $\sim x$. Its form 
is valid irrespective of the underlying microscopic dynamics, which 
certainly does not contain any such term.  In a sense, $x$ only measures the 
``entropic distance" to equilibrium. 

The same is true for the heat diffusion equation (in a stationary medium), 
where the only current $\sim\nabla T$ is again entropic and does not reflect 
an underlying microscopic dynamics: The thermodynamic force $\nabla T$ 
vanishes in equilibrium, same  as $x$, and is again a measure of the 
distance from it. So both types of macroscopic dynamics are in fact quite 
similar in their construction, with the only difference that the variable 
carrying out diffusion macroscopically is a conserved quantity. In the cases 
$\pi_{ij}+\pi^D_{ij}$, ${\bf H+H}^D$, ${\bf E+E}^D$, the respective first 
term does reflect the microscopic dynamics, while the second term is 
entropic. They measure the distance from equilibrium and nudge the 
respective field towards it. 

Summarizing, we conclude that since ${\bf E}^M$ and ${\bf H}^M$ are not 
simply the spatial average of ${\bf e}$ and ${\bf h}$, there is no good 
reason to rule out the dissipative fields ${\bf E}^D$ and  ${\bf H}^D$. A corollary
result is the difference between the 
stationary solution, $\nabla\times\bf{E}^M$, $\nabla\times\bf{H}^M=0$, and 
the equilibrium configuration $\nabla\times\bf{E}$, $\nabla\times\bf{H}=0$. 
The same must hold in the linear response theory, but does not in its usual 
version, as we shall find out in the next section. 

\section{Linear Response Revisited}\label{linres}
 Let us recapitulate the usual derivation of the macroscopic Maxwell 
equations, due to Lorentz. Start from the microscopic Maxwell equations 
containing only ${\bf b}$ and ${\bf e}$, divide the charge into 
$\langle\varrho_e\rangle$, $\nabla\cdot{\bf p}$, the current into 
$\langle{\bf j_e}\rangle$, $\partial_t{\bf p}$, $\nabla\times{\bf m}$, and 
define ${\bf h}={\bf b}-{\bf m}$, ${\bf d}={\bf e}+{\bf p}$ to eliminate 
${\bf p}$, ${\bf  m}$ while preserving the structure of the original 
equations. These simple, identical algebraic manipulations already yield 
the structure of the macroscopic Maxwell equations, albeit in terms of ${\bf 
b}, {\bf h}, {\bf  d}, {\bf e}$. Therefore, these equations are still 
detailed, reversible,  and completely equivalent to the starting point. A 
follow-up spatial averaging is easily executed, as the Maxwell equations are 
linear, and substitute $\langle{\bf b}\rangle, \langle{\bf h}\rangle, 
\langle{\bf  d}\rangle, \langle{\bf e}\rangle$ for ${\bf b}, {\bf h}, {\bf  
d}, {\bf e}$. Now these equations are of low resolution, yet they have 
retained the time reversal symmetry. 

The next step, deceptively simple, is the crucial one. It introduces 
irreversibilities and takes us across the Rubicon into coarse-grained, 
macroscopic physics --- although this step is usually considered the input 
of material properties, extrinsic to the Maxwell equations proper. We 
identify the variables, ${\bf B}\equiv\langle{\bf b}\rangle$, ${\bf  
D}\equiv\langle{\bf  d}\rangle$, and (instead of going through the averaging 
over ever increasing ensembles) substitute $\langle{\bf h}\rangle, 
\langle{\bf e}\rangle$ with ${\bf H}^M, {\bf E}^M$, and take the latter as 
functions of ${\bf B}, {\bf  D}$ and their temporal derivatives.  Retaining 
only first order derivatives, we have 

\begin{eqnarray}\label{18}
{\bf E}^M&=&{\bf D}/\bar\varepsilon+(\beta/\varepsilon_0)\dot{\bf 
D},\nonumber\\
{\bf H}^M&=&{\bf B}/ \bar\mu +(\alpha/\mu_0) \dot{\bf B}, 
\end{eqnarray}
or in Fourier space,  
\begin{eqnarray}\label{19}
{\bf D}&=& \bar\varepsilon{\bf E}^M/ 
(1-i\omega\beta\bar\varepsilon/\varepsilon_0),\nonumber\\
{\bf B}&=& \bar\mu{\bf H}^M/ (1-i\omega\alpha\bar\mu/\mu_0),
\end{eqnarray}
where $\bar\varepsilon, \bar\mu, \alpha, \beta$ are positive coefficients 
chosen to coincide with the hydrodynamic notation. Note especially the lack 
of time reversal symmetry of Eqs(\ref{18}). 

Two points here need amplification. First, it would not have been correct 
to take ${\bf D}$ as a function of  ${\bf E}^M$ and its derivatives: 
Generally, starting from a temporally nonlocal relationship between ${\bf 
D}$ and ${\bf E}^M$, both seem possible, and one would need microscopic 
details to decide which is realized in a specific case. Furthermore, since 
the microscopic field $\bf d$ contains the polarization $\bf p$, it is 
considered an auxiliary field, and preferred by many to depend on the true 
field $\bf e$ --- to lowest order in $\omega$ in the form 

\begin{equation}\label{20}
{\bf D}= \bar\varepsilon ({\bf E}^M- \beta\bar\varepsilon\dot{\bf 
E}^M/\varepsilon_0),
\end{equation}
written such that $\varepsilon=D/E^M$ remains unchanged from Eq(\ref{19}) 
in the given order. However, this formula is unacceptable on general grounds, 
irrespective of the microscopics. Assume homogeneity and stationarity in 
Eqs(\ref{14}), $\nabla\times{\bf E}^M$, $\nabla\times{\bf H}^M$,  $\dot{\bf 
D}$, $\dot{\bf B}=0$, and find Eq(\ref{20}) producing a run-away solution 
$\dot E^M \sim \exp(t/ \beta\bar\varepsilon)$. (One must not change the 
minus sign in Eq(\ref{20}) to render the solution decaying, as this would 
result in self-amplifying electromagnetic waves.)

The second point concerns the lack of spatial nonlocality, or 
why the permeabilities $\varepsilon$ and $\mu$ have not been taken as 
functions of the wave vector $q$. Usually, the answer entails a discussion 
of correlation lengths, which (for simplicity) takes place in an infinite 
medium. And the result is that spatial nonlocality becomes important only 
at microscopically small scales, so is usually negligible --- except perhaps 
in systems with fast moving charge carriers, such as dilute plasmas. Let us, 
however, consult  the hydrodynamic theory, the concept and considerations of 
which naturally include boundaries. It is in fact quite unambiguous on this 
point, and it must be completely equivalent to the linear response theory in the 
parameter space where both the frequency and the field strength are low, and 
where the two stripes, A and B of Fig~1, overlap. Assuming constant 
temperature, we may rewrite Eqs(\ref{16a},\ref{16}) as 

\begin{eqnarray}\label{21}
&{\bf E}^D= (\beta/\varepsilon_0){\bf\dot 
D}&+\lambda^2\nabla\times\nabla\times{\bf E},\\\label{22}
&{\bf H}^D=(\alpha/\mu_0){\bf\dot B}&+\lambda^2\nabla\times\nabla\times{\bf 
H}, \end{eqnarray}
where
\begin{equation}
\lambda=c\sqrt{\alpha\beta}.
\label{22a}\end{equation}
If either $\alpha$ or $\beta$ is very small, the two terms $\sim\lambda^2$ 
may be neglected. Then these two equations -- via Eqs(\ref{7a}) and 
(\ref{15}) -- reduce to the usual linear response expressions, 
Eqs(\ref{18},\ref{19}). However, there are also cases in which $\lambda$ 
cannot be neglected: Being proportional to the relaxation time of 
magnetization and polarization, cf Eq(\ref{17}), $\alpha$ and $\beta$ vary 
greatly, from $\beta\approx10^{-15}$s for transparent dielectrics, to 
$\alpha\approx10^{-5}$s for colloidal magnetic liquids (Rosensweig 1985, 
Shliomis 1974); while water (or any other matrix liquid with a strong, 
permanent molecular dipole moment) is in 
the middle range, $\beta\simeq10^{-9}$s. So a water-based ferrofluid should 
have a colossal $\lambda\simeq 3\rm x10^3$cm. 

There is a statement about dielectric media, found in every textbook on 
electromagnetism: If the electric and magnetic field are static, they are 
longitudinal and decoupled from each other. It is true if the usual linear 
response theory, Eqs(\ref{18},\ref{19}), holds: Setting ${\bf \dot D}$, 
${\bf \dot B}=0$, we have $\nabla\times{\bf E},\  \nabla\times{\bf H}=0$, 
and the only stationary solution are indeed longitudinal, decoupled and in 
equilibrium. But it is not true in the general case, as the more complete 
version of the linear response theory, Eqs(\ref{21},\ref{22}), or simpler, 
the hydrodynamic theory, Eqs(\ref{14},\ref{15},\ref{16}), entertains a 
stationary solution that is off-equilibrium and of the form 

\begin{eqnarray}\label{23}
E_x&=&({\cal E_+}e^{z/\lambda}+{\cal 
E_-}e^{-z/\lambda})/\sqrt{\varepsilon_0},\\ \label{23a}
H_y&=&({\cal E_+}e^{z/\lambda}-{\cal 
E_-}e^{-z/\lambda})\sqrt{\alpha/(\beta\mu_0)},
\end{eqnarray}
where $\cal E_\pm$ are constant amplitudes. These electric and magnetic 
field are coupled and transverse, they start off from the boundary and relax 
exponentially into the bulk. Since the presence of a boundary makes itself 
felt over the distances of $\lambda$, a dielectric ferrofluid is indeed a 
dense and strongly interacting system that entertains a macroscopically 
large spatial nonlocality. It would be interesting to detect these fields, 
and fortunately, this does not appear to be difficult. 

{\bf[}Take a slab of dielectric ferrofluid, with a width $L\ll\lambda$, say 
1cm.  Expose this liquid to an oscillating electric or magnetic field, 
tangential to the slab, of the frequency $\omega\ll c/L$, and measure the 
internal field. Conventionally, we expect the result to be a uniform 
internal field, $E$ or $H$, that oscillates in phase with the external one 
and has the same magnitude, so $D$ and $B$ display a phase lag  
$\sim\varepsilon'', \mu''$. The hydrodynamic consideration includes the 
stationary solution above, and the result depends on the type of the 
containing plates: If they are nonconducting, the phase lag has the same 
magnitude but an opposite sign; if they are conducting, the internal 
electric field should be drastically reduced (Liu 1997).{\bf]}

\section{Electromagnetic Forces}\label{force}
Given the input of the thermodynamic theory, the basic form of the Maxwell 
equations and the relevant conservation laws, the derivation of the complete 
hydrodynamic theory is an exercise in cogent deduction and algebraic 
manipulation (Khalatnikov 1965, Henjes and Liu 1993). One of the 
main results is the expression for the total stress tensor, $\Pi_{ij}= 
\pi_{ij}+ \pi^D_{ij}$: The first is the Lorentz-Galilean boosted Maxwell 
tensor, the second is the dissipative, off-equilibrium contribution, of the 
usual form and $\sim v_{ij}$ if the liquid is isotropic and the external 
field weak. {\large\bf[}Otherwise, it will contain terms 
$\sim\nabla\times{\bf E}$, $\nabla\times{\bf H}$, cf Liu (1994).{\large\bf]} 
Inserting the explicit expression for $\Pi_{ij}$ into Eq(\ref{11}), and combining 
it with and Eqs(\ref{14}), we obtain a transparent form of the momentum 
conservation, 

\begin{eqnarray}
 &\rho(\textstyle{\frac{\partial}{\partial t}}+v_j\nabla_j)[v_i+({\bf 
E\times H}/c^2-{\bf D\times 
 B})_i/\rho]\nonumber\\ 
&+s\nabla_iT+\rho_1\nabla_i\mu_1+ 
\rho_2\nabla_i\mu_2+g_j\nabla_iv_j\nonumber\\ 
&=\left(\langle\rho^e\rangle{\bf E}+\langle{\bf j^e\rangle\times B} +{\bf 
f^D}\right)_i-\nabla_j\pi^D_{ij}.\label{vdot} \end{eqnarray}The first line 
contains both the acceleration $\rho\mbox{\boldmath$\dot v$}$ and the 
Abraham-force $\frac{\partial}{\partial t}({\bf E\times H}/c^2- {\bf D\times 
B})$. (The latter is a small quantity if the electromagnetic wave length 
of a given frequency is large compared to the experimental dimension, as is 
usual for hydrodynamic frequencies.) The four terms of the next line are the 
proper generalization of the pressure gradient and include the reactive 
ponderomotive forces. They are valid for a two-component medium (such as a 
solution), with $\rho_1$, $\rho_2$ denoting the respective density, 
and $\rho=\rho_1+\rho_2$. Given 
linear constitutive relations, assuming that $\bar\varepsilon-\varepsilon_0$, 
$\bar\mu-\mu_0$ are proportional to one of the densities but independent of the 
other, and neglecting contributions $\sim{\cal O}(v/c)$, they reduce to the 
usual Kelvin force,   

\[\nabla P-\textstyle{\frac{1}{2}}(\bar\varepsilon-\varepsilon_0)\nabla 
E^2-\textstyle{\frac{1}{2}}(\bar\mu-\mu_0)\nabla H^2, \]
where $P$ is the pressure in the absence of fields. The third line, finally, contains the 
dissipative stress tensor $\pi^D_{ij}$, the Lorentz force --- in terms of $E$ rather than 
$E^M$, in contrast to Eq(\ref{5}) --- and the novel dissipative ponderomotive force, 

\begin{equation}
{\bf f^D}={\bf B\times\nabla\times H^D+D\times\nabla\times E^D}. 
\end{equation}
It is odd under time reversal, same as $\pi^D_{ij}$, and accounts for forces 
that arise because the polarization and magnetization are not quite in 
equilibrium. 

As we shall soon realize, terms in ${\bf f^D}$ of linear order in $v/c$ may 
be just as  important as those of zeroth order. They have been neglected 
above to keep the arguments and display simple, but are easily retrieved: 
The fields in the Euler Eqs(\ref{9}), and hence in the dissipative fields, 
are those of the local rest frame, 

\begin{eqnarray}
{\bf H}^D&=&-(\alpha/\mu_0)\nabla\times{\bf E^0},\quad 
{\bf E}^D=(\beta/\varepsilon_0)\nabla\times{\bf H^0},\\
{\bf E^0}&=&{\bf E}+ \mbox{ \boldmath$v$} \times{\bf B},\quad 
{\bf H^0}= {\bf H}-\mbox{\boldmath$v$} \times {\bf D}.
\end{eqnarray}
This is plausible: The question whether equilibrium is established in a 
system cannot depend on the observer's choice of the inertial frame. 
{\large\bf[}The derivation consists of generalizing the results of 
Eqs(\ref{8},\ref{9},\ref{16}) to systems with  finite velocities, where a 
pragmatic combination of Galilean and first order Lorentz transformation is 
employed. A fully covariant theory (Kost\"adt and Liu 1998) was also derived to 
make sure that the additional terms are indeed negligible in usual circumstances 
(Symalla and Liu 1998).{\large\bf]}

Assuming for simplicity that $\beta$ (and hence $E^D$) is negligible, and 
confining our considerations to incompressible systems of simple geometry 
--- a sphere or a slab such that the internal field are uniform --- we may 
rewrite $H^D$ as 

\begin{equation}\label{24}
{\bf H^D}=(\alpha/\mu_0)[{\bf \dot B}-({\bf 
B}\cdot\nabla)\mbox{\boldmath$v$}], \end{equation}
where the second term, of order $v/c$, is a result of the fact that ${\bf H^D}$ is given in 
terms of the restframe field ${\bf E^0}$,  rather than ${\bf E}$. 

 First, consider a solid magnetic sphere, $\mbox{\boldmath$v$} = 
\Omega\times \mbox{ \boldmath$r$} $, and expose it to a rotating magnetic 
field of the frequency $\Omega_B$. If there are no further perturbations, 
all forces except ${\bf f^D}$ vanish, and ${\bf H^D}=(\alpha/\mu_0)[( 
\mbox{\boldmath$\Omega_B- \Omega$}) \times{\bf B}]$.  The torque on the 
sphere,  after being partially integrated (over a surface slightly larger 
than the sphere), is $ \int dV (\mbox{\boldmath$r$}\times{\bf f^D}) = \int 
dV ({\bf B\times H^D})$. With $\Theta$ denoting the sphere's moment of 
inertia, Eq(\ref{11}) therefore reduces to, 

\begin{equation}\label{25}
\mbox{\boldmath$\dot\Omega$}=(\alpha V/\Theta\varepsilon_0)\ {\bf 
B}\times[( \mbox{\boldmath$\Omega_B-\Omega$})\times{\bf B}]. \end{equation}
After an initial surprise, it is reassuring to see how these 
two terms of successive orders in $v/c$ cancel each other if the magnetic 
field and the sphere co-rotate -- as they must to conserve total angular momentum. 
(A curiously similar effecr at linearly polarized magnetic field was observed by 
Gazeau et al, 1997)

Next, consider a slab of ferrofluid sustaining a shear flow $\nabla_nv_t$, 
where the subscripts denote the normal and tangential direction with respect 
to the slab. If $\dot B=0$, only the second term of  Eq(\ref{24}) remains, 
which represents a correction to the viscosity $\eta$, 

\begin{equation}\label{26}
\rho\dot v_t=\eta_e\nabla^2_n v_t,\quad\eta_e=\eta+(\alpha/\mu_0) 
B^2\cos^2\varphi, \end{equation}
where $\varphi$ denotes the angle between $B$ and the slab normal. {\bf[}In 
the presence of a magnetic field, the viscosity $\eta$ itself is a function 
of $B$ and its orientation. Therefore, the information about this 
contribution of $\alpha$ is not easily extracted experimentally.{\bf]}    

If the field oscillates in time, the first term in Eq(\ref{24}) becomes 
important, $\rho\dot v_t -\eta_e \nabla^2_nv_t  = B_n\dot B_t  
\nabla_n\alpha/\mu_0$. Because $\alpha$ (as a function of temperature and 
density) is constant within the ferrofluid,  the right side is nonvanishing 
only at the surface, where $\alpha$ jumps discontinuously to zero. So this 
force is best accounted for with boundary conditions. Solving $\rho\dot v_t 
= \eta_e\nabla^2_nv_t$, the boundary conditions are 

\begin{equation}\label{27}
\eta_e\nabla_n v_t= (\alpha/\mu_0) B_n \dot B_t,\qquad v_t=0,
\end{equation}
for a free and sticking surface, respectively. {\large\bf[}The free 
condition is obtained by setting the off-diagonal stress tensor $\Pi_{tn}$ 
to zero, while heeding the fact that $H^M_t$, rather than $H_t$, is 
continuous across the interface.{\large\bf]} 

Expose a slab of ferrofluid with its free surface facing upward to a static 
normal and an oscillating tangential field. If the frequency is low enough,  
we may neglect $\rho\dot v_t$, and the gradient given by Eq(\ref{27}) is 
constant throughout the width $L$ of the slab. Since $v_t=0$ at the bottom,  
the averaged velocity is $\langle v_t\rangle=(\alpha/\mu_0) B_n\dot 
B_tL/2\eta_e$. 

 Now forget the gravity, bend this slab into a ring with the free surface 
facing inward, and expose this construct to a $B$-field rotating in the 
plane of the ring. Take the momentary orientation of the field as $\hat x$, 
then a counter-clockwise rotation has $\dot B$ along $\hat y$. In the two 
sections of the ring perpendicular to $\hat x$, the gradient of the 
velocity, according to Eq(\ref{27}), is positive. Because the velocity must 
vanish at the outer rim of the two sections, it is negative in the right 
section, and positive in the left. They combine to yield a clockwise 
circular flow, opposite to the external field. 

 In an open, round vessel, the free surface of the ferrofluid (facing 
upward) curves up at the wall, as wetting fluids do. And the capillary 
region is very similar to such a sheet of circular flow, and should rotate 
opposite to the external field --- while the bulk of the fluid below 
rotates, more or less, with this region (Rosensweig et al 1990). 

All these effects follow cogently from the identification of  
$\nabla\times{\bf E^0}$ as the magnetic  thermodynamic force. They depend on 
only one parameter $\alpha$ that has a clear-cut physical significance and a 
definite value. In addition, there is a one-to-one correspondence to the 
analogous effects stemming from the dissipative, ponderomotive electric 
force, ${\bf D\times\nabla\times E^D}$.

\end{document}